\documentclass[pre,aps,floatfix,twocolumn,showpacs,preprintnumbers,superscriptaddress]{revtex4-1}
\usepackage{amsmath}
\usepackage{amssymb}
\usepackage{graphicx}
\usepackage{color}

\newcommand{\vp}{\varphi}
\newcommand{\w}{\omega}

\newcommand{\e}{\varepsilon}

\begin{document}
\title{Reconstructing networks of pulse-coupled oscillators from spike trains}
\author{Rok Cestnik}
\author{Michael Rosenblum}
\affiliation{Department of Physics and Astronomy, University of Potsdam,
 Karl-Liebknecht-Str. 24/25, D-14476 Potsdam-Golm, Germany}
\date{\today}

\begin{abstract}
We present an approach for reconstructing networks of pulse-coupled 
neuron-like oscillators from passive observation of pulse trains of all nodes. 
It is assumed that units are described by their phase response curves 
and that their phases are instantaneously reset by incoming pulses.
Using an iterative procedure, we recover the properties of all nodes, 
namely their phase response curves and natural frequencies, as well as strengths of
all directed connections. 
\end{abstract}
\maketitle

\section{Introduction}

Reconstruction of a network structure from observations is an important problem 
relevant for many different areas such as neuroscience 
\cite{Sporns-05, *Riera-05, *Beckmann-05,*Bullmore-Sporns-09,%
*Rubinov-Sporns-10,*Friston-11,*Chicharro-Andrzejak-Ledberg-11,%
*Lehnertz-11,*Boly_et_al-12,*Sporns-13}, 
physiology \cite{Mrowka_et_al-03,*Musizza_et_al-07,*Kralemann_et_al-13}, 
climatology \cite{Wang_et_al-12,*Sharma_et_al-12}, 
genetics \cite{Gardner-04,*Zhao-08}, 
ecology \cite{Berlow-04,*Emmerson-04,*Sugihara26102012,*Gray-15}, etc.
A group of established reconstruction techniques relies 
on analysis of the system's response to a specially designed 
perturbation, i.e. on invasive measurements 
\cite{Timme-07,*Yu-Parlitz-08,*Yu-11,*Levnajic-Pikovsky-11}. 
However, often invasive measurement is not an option, e.g. 
in problems related to climatology, physiological studies, 
and medical diagnostics.
In such cases one is restricted to analysis of 
observations of the free-running system.

Roughly speaking, there are two approaches to the problem. 
The first one does not imply any assumptions about the dynamics of the nodes 
and properties of the links and relies on different statistical and 
information-theoretical techniques for quantification of all connections 
\cite{Aertsen-85, *Schreiber-00, *Palus-Stefanovska-03,*PhysRevLett.99.204101,%
*PhysRevLett.100.158101,*PhysRevLett.103.238701,*Chicharro-09, *Andrzejak-11,%
*PhysRevE.83.051112,*PhysRevLett.108.258701,*Mishchencko-11,%
*Battaglia-Witt-Wolf-Geisel-12,*Kugiumtzis-13,*Tirabassi-15}. 
In the second, model-based approach, some properties of the nodes 
(e.g. existence of a stable limit cycle) and of the links 
(e.g. weakness of coupling) are assumed to be known 
\cite{Rosenblum-Pikovsky-01,*Rosenblum_et_al-02,*Kralemann_et_al-07,%
*Kralemann_et_al-08,Galan-Ermentrout-Urban-05,Kralemann-Pikovsky-Rosenblum-11,%
Stevenson-08,*Valdes-Sosa-11}.
In the present work we follow and extend the model-based approach. 
The main assumption is that the networks can be modeled by coupled 
limit cycle oscillators \cite{Pikovsky-Rosenblum-Kurths-01,Izhikevich-07}.
In this way we follow our previous studies, where we have reconstructed the 
connectivity of a weakly coupled network of noisy limit-cycle 
or weakly chaotic oscillators for the case when the measurements allow for 
the determination of instantaneous phases 
\cite{Kralemann-Pikovsky-Rosenblum-11,Kralemann-Pikovsky-Rosenblum-14}, see also
\cite{Penny-Litvak-Fuentemilla-Duzel-Friston-09,*Cadieu2010,%
PhysRevLett.109.024101,*Rings-Lehnertz-16}. 

In this paper we address the case when the signals
are spiky, namely, 
that the measurements between the spiking
events are dominated by noise
and only determination of the times of spikes is reliable.
Hence, the data we analyze are spike trains and 
estimation of time-continuous phase is not feasible. 
Next, we assume that effect of a chosen unit on the rest of the network
is restricted to the time instant when the unit generates a spike. 
Thus, we use the model of pulse-coupled neuron-like oscillators 
\cite{Mirollo-Strogatz-90,*Ernst-Pawelzik-Geisel-95,*vanVreeswijk-96,%
*Gerstner-96,*Mohanty-Politi-06,*Makarov-05,Memmesheimer-Timme-06,%
*Kinzel-08,*Patnaik-08, *VanBussel-11,*Barranca-16}.
Assuming that the outputs of all nodes are known and that the 
coupling between the elements is sufficiently weak to justify the 
phase dynamics description, we recover the connectivity of the network
and properties of all its nodes.

The paper is organized as follows. In Section~\ref{sec:model} we describe 
in details the model and summarize all the assumptions. 
In Section~\ref{sec:techn} we introduce our technique and in 
Section~\ref{sec:num} we present the results of numerical studies.
Section~\ref{sec:concl} presents discussion of the results.

\section{The model}
\label{sec:model} 
Our basic model for the network's node is a limit cycle oscillator
which issues a spike when its phase $\vp$ achieves $2\pi$. 
(We consider the phases wrapped to the $[0,2\pi)$ interval, 
i.e. after the spike generation the phase of the unit is reset to zero).
This spike affects all other units of the network according to the
strength of the corresponding out-coming connections. 
Let the size of the network be $N$ and let the connectivity be described 
by an $N\times N$ coupling matrix ${\cal E}$, whose elements $\e_{ij}$
quantify the strength of the coupling from unit $j$ to unit $i$.
Between the spiking events, phases of all units obey $\dot\vp_i=\w_i$,
where $\w_i$ are frequencies.
If unit $i$ receives a spike from oscillator $j$, then it reacts
to the stimulus according to its so-called phase response curve (PRC), 
$Z_i(\vp)$ \cite{Winfree-80,*Glass-Mackey-88,Canavier-06}. 
This means that the phase of the stimulated unit is 
instantaneously reset, $\vp_i\to \vp_i+\e_{ij}Z_i(\vp_i)$. 

Notice that oscillators are  generally non-identical: they have different
frequencies and different PRCs.
However, we assume that response of the unit $i$ to the stimuli 
from different units is described by the same PRC $Z_i$.
Furthermore, we assume that PRCs are continuous.
Next, the coupling is taken to be bidirectional but generally
asymmetric, i.e. $\e_{ji}\ne \e_{ij}$, and there is no self-action, i.e. 
$\e_{ii}=0$.

\begin{figure}[h!]
\centering
\includegraphics[width=0.98\columnwidth]{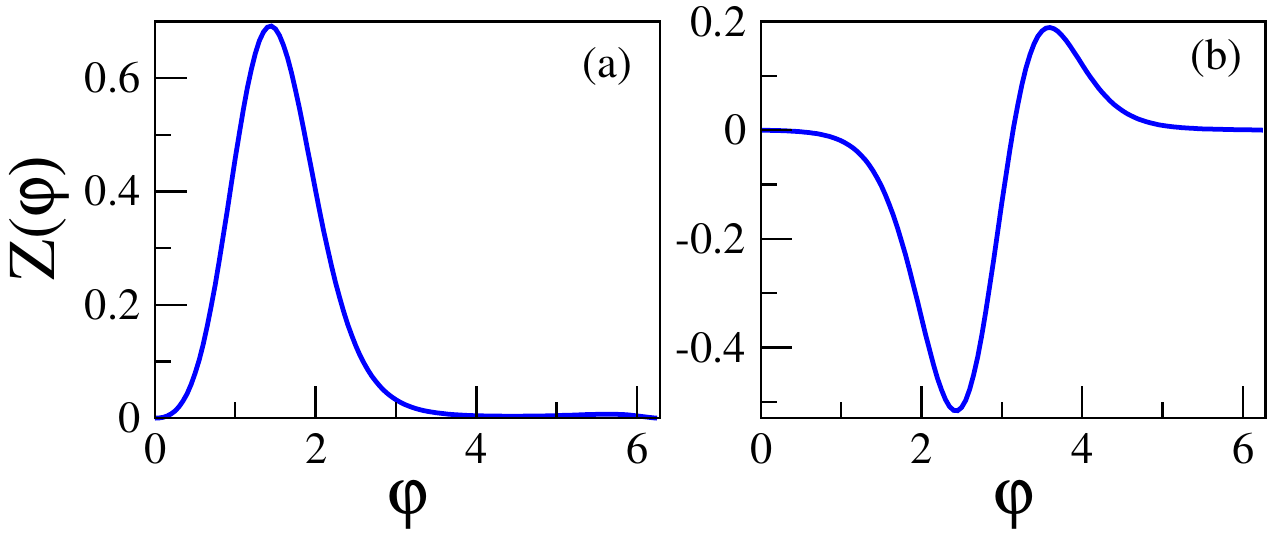}
\caption{Model phase response curves of type I (a) and type II (b).}
\label{modprc}
\end{figure}

In neuronal modeling one commonly identifies two types of PRCs: 
if spikes always shorten the period of the stimulated unit, then the PRC is 
classified as type I.
Otherwise, if depending on the phase of the stimulation, the period can 
be either shortened or prolonged, then the PRC is classified as type II
\cite{Hansel-Mato-Meunier-95,Canavier-06}.  
We model the type I PRC as
\begin{equation}
Z(\vp)=(1-\cos(\vp)) \exp{\left(3[\cos(\vp-\vp_0)-1]\right )}\;,
\label{prc1}
\end{equation}
and the type II PRC as
\begin{equation}
Z(\vp)=-\sin(\vp) \exp{\left(3[\cos(\vp-\vp_0)-1]\right )}\;,
\label{prc2}
\end{equation}
where the parameter values are $\vp_0=\pi/3$ and $\vp_0=0.9\pi$, respectively.
The plots of these curves are shown in Fig.~\ref{modprc}. 

Using this model we generate $N$ point processes (spike trains) and then use them 
for network reconstruction, where we estimate the coupling matrix ${\cal E}$, 
PRCs $Z_i(\vp)$, and frequencies $\w_i$ of all elements, as discussed in the 
next section.

\section{The technique}
\label{sec:techn}
For each node we reconstruct its properties as well as strength of all 
incoming connections. For definiteness, we always determine these quantities 
for the first node; the procedure then shall be repeated for all other units.
Thus, we recover $\e_{1j}$, $Z_1$, and $\w_1$; for simplicity of presentation, 
in the following we omit the subscript $1$.

We solve the reconstruction problem by iterations. 
First, since we do not have any \textit{a priori} knowledge of the system, 
we assign some values to the coupling coefficients (we discuss several 
option of how this can be done) and use them in order to obtain a first  
estimate of the PRC.
The knowledge of the latter allows for an improved estimation 
of the network connectivity, which is then in turn used to obtain a better 
approximation of the PRC, and so on. We demonstrate that the procedure converges
quite fast.

\subsection{Notations and phase equations}
\begin{figure}[h!]
\centering
\includegraphics[width=0.8\columnwidth]{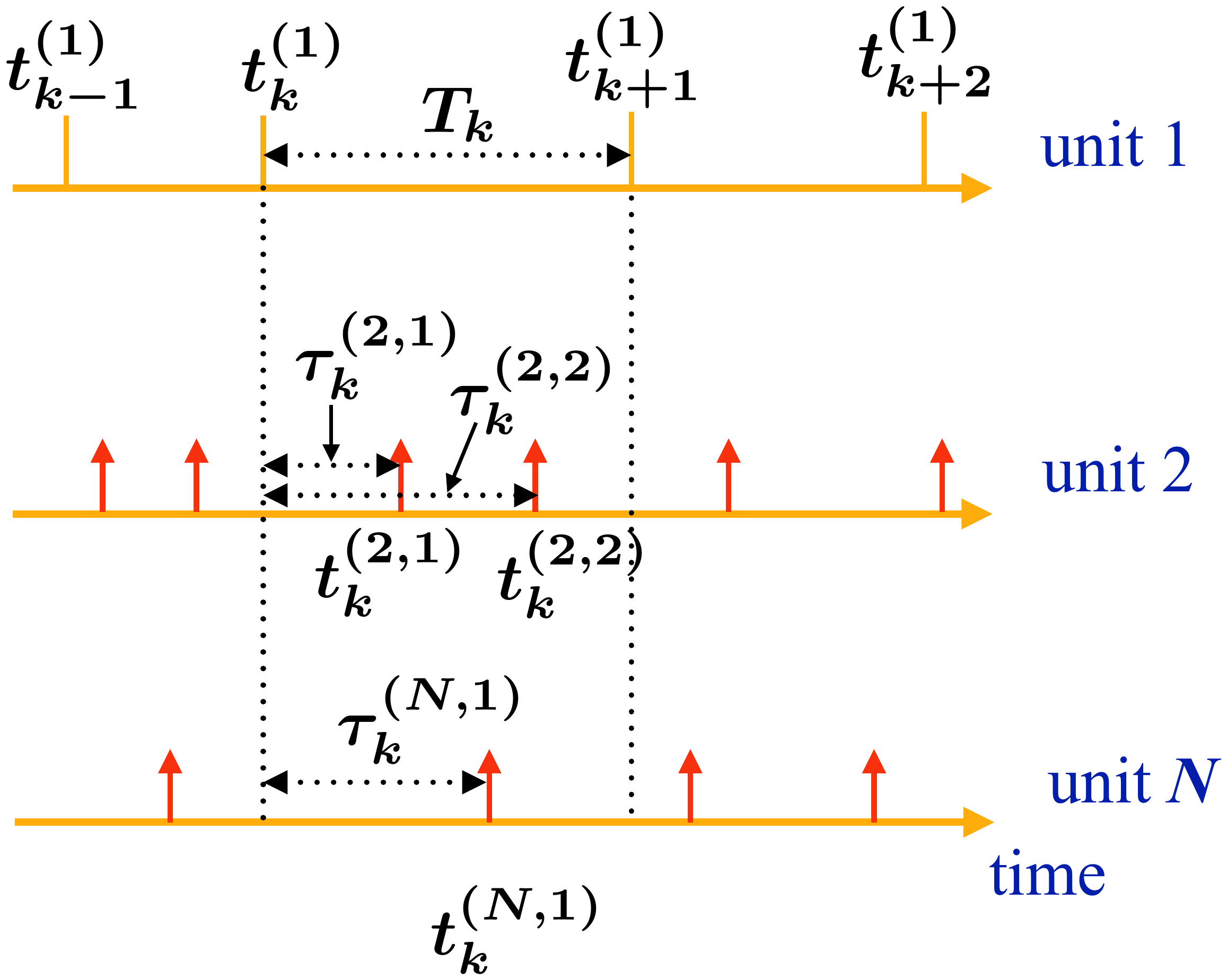}
\caption{Illustration of notations used. $T_k$ is an inter-spike
interval of the driven unit. $\tau_k^{(i,l)}$ is the time 
(relative to the beginning of the interval $T_k$) when the 
spike number $l$ from the unit number $i$ arrives.
}
\label{notations}
\end{figure}

Let the pulse train of the first oscillator
contain $M+1$ spikes at times $t_k^{(1)}$, so that we have $M$ inter-spike 
intervals $T_k=t_{k+1}^{(1)}-t_k^{(1)}$.
In the following we treat each interval separately.
Suppose that within the inter-spike interval 
$T_k$ the first
unit receives $n$ stimuli from the unit $i$, we denote this number as $n_k(i)$.
These stimuli appear at instants of time $t_k^{(i,l)}$, $l=1,\ldots,n_k(i)$.
The times relative to the beginning of the interval are denoted as 
$\tau_k^{(i,l)}=t_k^{(i,l)}-t_k^{(1)}$; see Fig.~\ref{notations} 
for illustration.
Respective phases of the first unit are denoted as 
$\vp(t_k^{(1)}+\tau_k^{(i,l)})=\vp_k^{(i,l)}$.

The phase increase within each inter-spike interval is
\begin{equation}
\w T_k+\sum_{i=2}^N\e_i\sum_{l=1}^{n_k(i)}Z(\vp_k^{(i,l)}) = 2\pi\;,
\label{eq1}
\end{equation}
where the first term reflects the autonomous dynamics, whereas the second term
describes the effect of pulse coupling.
$M$ inter-spike intervals yield a system of $M$ Eqs.~(\ref{eq1}) for unknown
coupling coefficients $\e_i$, frequency $\w$, and the PRC $Z$ of the 
driven unit. 

Assume for the moment that the coupling coefficients $\e_i$ are given. 
Then, representing the unknown $Z(\vp)$ as a finite Fourier series
of order $N_F$, we obtain from Eqs.~(\ref{eq1}) a system of linear 
equations for $2N_F+1$ Fourier coefficients and the unknown frequency $\w$.
For a long time series, $M > 2N_F+2$, 
this is an over-determined system which can be solved, e.g. by a 
least-mean-square fit or by singular value decomposition,
see \cite{Pikovsky-16}. On the other hand, if PRC is given, 
we again obtain a solvable linear system for $N-1$ coupling 
coefficients $\e_i$ and frequency $\w$~\footnote{ 
When neither PRC nor $\e_i$ are known, 
Eqs.~(\ref{eq1}) represent a nonlinear system with respect to $N+2N_F+2$ unknowns. 
Alternatively, one could consider products of $\e_i$ and the Fourier coefficients 
as unknowns and end up with a linear, but rather large system of $N(2N_F+2)$ 
unknowns.}.
Thus, having an initial estimate for either PRC or coupling 
coefficients  (practically we use the later option) we can try to solve 
Eqs.~(\ref{eq1}) by iterations.

\subsection{First iteration}

The phases within each inter-spike interval vary from zero to $2\pi$.
For the first iteration we take the simplest approximation, i.e. we  
compute the phases as growing proportionally to time. 
Thus, when a spike at $\tau_k^{(i,l)}$ arrives, the phase of the first
unit is taken as 
\begin{equation}
\vp_k^{(i,l)} \approx 2\pi\tau_k^{(i,l)}/T_k  \;.
\label{eq3}
\end{equation}
Since in this approximation we neglect the phase resets, 
$\vp\to \vp+\e_{i}Z(\vp)$, the errors of such a phase estimation 
are of the order of $\e_i\Vert Z\Vert$, where $\Vert\cdot\Vert$ means 
norm of the function, and accumulate with the number of the incoming 
spikes.

Next, we have to choose some initial values for the coupling coefficients
$\e_i$. There are several options how to do this. First, we can exploit 
the simple idea that if there is no connection to the first unit from 
the unit $i$, then $T_k$ cannot depend on the phase when the spikes
from this unit appear, i.e. there shall be no dependence of $T_k$ on 
$\vp_k^{(i,1)}$. On the other hand, if this connection exists, the 
dependence $T_k$ on $\vp_k^{(i,1)}$ shall be present as well; 
moreover, the larger $\e_i$, the stronger this dependence shall be.
As shown in Appendix~\ref{subsec:first_estimate}, this idea indeed
works well for long time series.
Two further variants are to assign initially same value to all 
$\e_i$ or take them randomly.

\subsection{Next iterations}
In the first approximation we compute the phases proportionally to 
time, see Eq.~(\ref{eq3}). 
If the coupling strength, $\e_i$, and parameters of the system, 
i.e. $\w$ and $Z$, are already estimated, 
then we can use this knowledge for a more precise estimation 
of the phases.
For illustration, suppose that within the inter-spike interval 
$T_k$ the first unit receives three stimuli at times 
$\tau_k^{(i,1)}<\tau_k^{(m,1)}<\tau_k^{(n,1)}$.
Then the phases at these three instances are computed as
\begin{align*}
\vp_k^{(i,1)}&=\w \tau_k^{(i,1)}\;,\\
\vp_k^{(m,1)}&=\w \tau_k^{(m,1)}+\e_iZ\left(\vp_k^{(i,1)}\right)\;,\\
\vp_k^{(n,1)}&=\w \tau_k^{(n,1)}+\e_iZ\left(\vp_k^{(i,1)}\right)
              +\e_mZ\left(\vp_k^{(m,1)}\right)\;.
\end{align*}
The phase at the end of the given inter-spike interval is
\[ 
\psi=\w T_k+\e_iZ(\vp_k^{(i,1)})+\e_mZ\left(\vp_k^{(m,1)}\right)
+\e_nZ\left(\vp_k^{(n,1)}\right)\;.
\]
By definition, this value should be equal to $2\pi$. However, since $\w$ and $Z$ are 
not exact, $\psi$ generally differs from $2\pi$. 
Therefore, we re-scale all phase estimates by the factor $2\pi/\psi$.  

Now, using the newly estimated phases and the estimation of the PRC
from the previous iteration, we can compute new values of the coupling 
coefficients $\e_i$, and then repeat the whole procedure. 
As we demonstrate below, these iterations converge quite quickly.

\section{Numerical tests}
\label{sec:num}

In this section we present the results of numerical testing of our reconstruction
algorithm.
For this goal we generate the networks with some randomly chosen parameters 
and then compare the reconstructed values with the true ones.
Namely, we consider networks of $N = 20$ oscillators, with natural frequencies 
taken from a uniform distribution between 1 and 2. 
Strength of network links is sampled from the positive half of a Gaussian distribution 
with zero mean and standard deviation $0.02$. 
We excluded from the consideration the networks where at least two units synchronized. 
The frequency of the first oscillator is set to 1, assuring that it is the slowest one 
(as discussed below, this is the most difficult case) 
and then we reconstruct its PRC $Z$, frequency $\w$, and strength of all 
incoming links $\e_i$, $i=2,\ldots,20$. We use ten iterations of the procedure 
described above.

\begin{figure}[htb!]
\centering
\includegraphics[width=1.01\columnwidth]{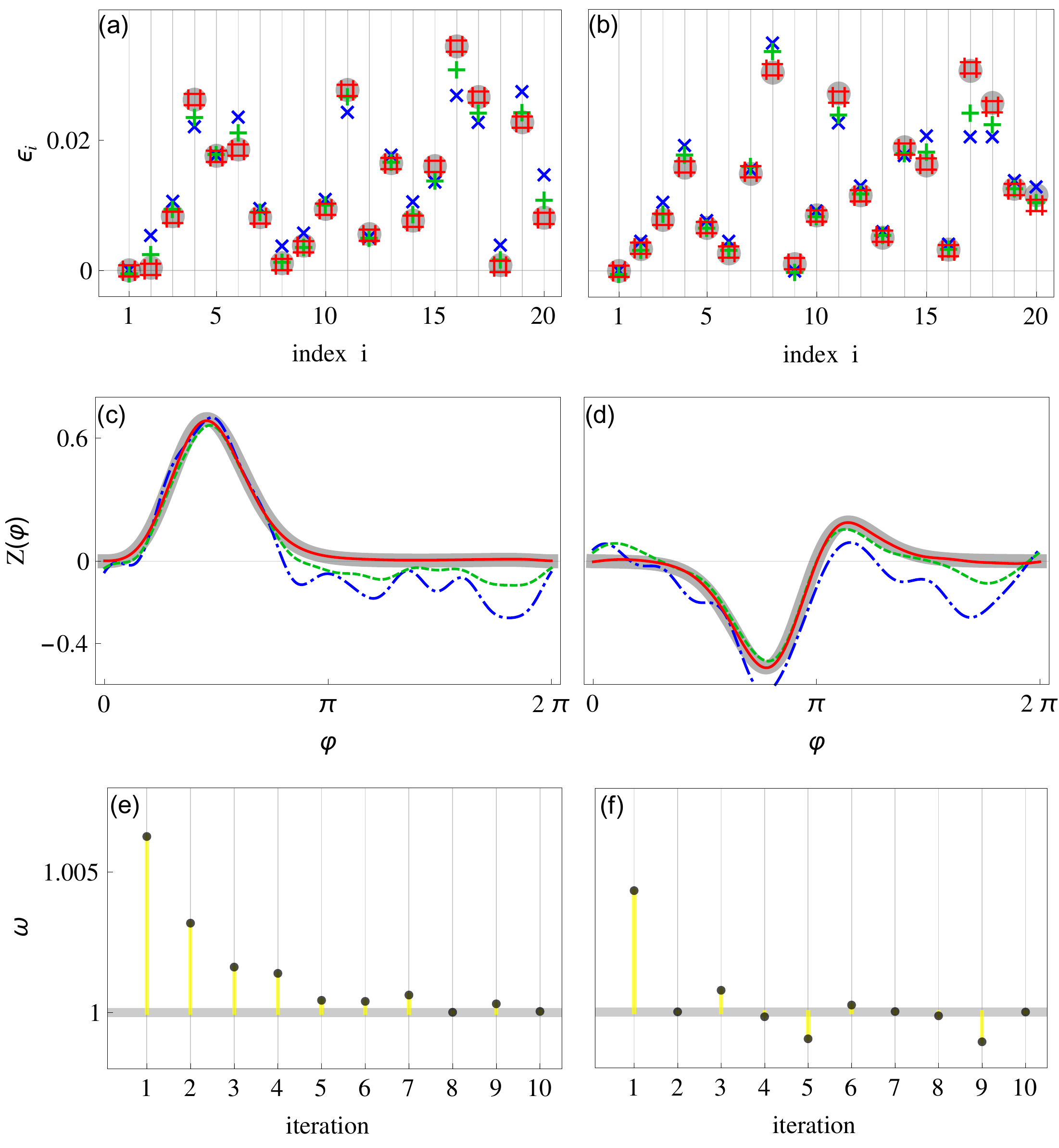}
\caption{(Color online) Reconstruction of a network of 20 units using $M = 200$ 
inter-spike intervals.
Panels (a,c,e) and (b,d,f) show the results for PRC type I and PRC type II,
respectively, see Eqs.~(\ref{prc1},\ref{prc2}). 
(a,b) Strength of the incoming connections to the first oscillator:
true values (gray disks) and values recovered after one, 2, and 10 iterations
(blue crosses, green pluses, and red hash-tag, respectively).
(c,d) True (wide gray curve) and reconstructed PRCs, after one, 2, and 10
iterations (blue dashed-dotted, green dashed, and red solid curves, respectively). 
(e,f) Estimated natural frequencies as functions of the iteration number; the true
value $\w=1$ is shown by horizontal gray line.
}
\label{fig:1run}
\end{figure}

Before presenting the results we recall that all equations contain 
only the products of $\e_i$ and $Z$. Hence, solutions $\e_i,Z$ and $c\e_i,Z/c$,
where $c$ is an arbitrary constant, are equivalent. 
The factor $c$ has no physical meaning by itself, but since we want to compare
the reconstructed values with the originally given, we have to fix it.
Quite arbitrarily, we do it by minimizing 
\[
\sum\limits_{i=2}^N \left[ \e_i^{\text{(t)}} - c \e_i^{\text{(r)}} \right]^2
\] 
where the superscripts $\text{(t)}$ and $\text{(r)}$ stand for true and reconstructed, 
respectively. This condition yields  
\begin{equation}
c = \sum\limits_{i=2}^N \e_i^{\text{(t)}} \e_i^{\text{(r)}} \bigg / 
\sum\limits_{i=2}^N \left[ \e_i^{\text{(t)}} \right]^2\;.
\label{eq:norm}
\end{equation}
Using this normalization, we show the results of a particular run 
in Fig.~\ref{fig:1run}; for this computation we took initially
$\e_i=1\,,\;\forall i$.

\begin{figure}[h!]
\centering
\includegraphics[width=1.01\columnwidth]{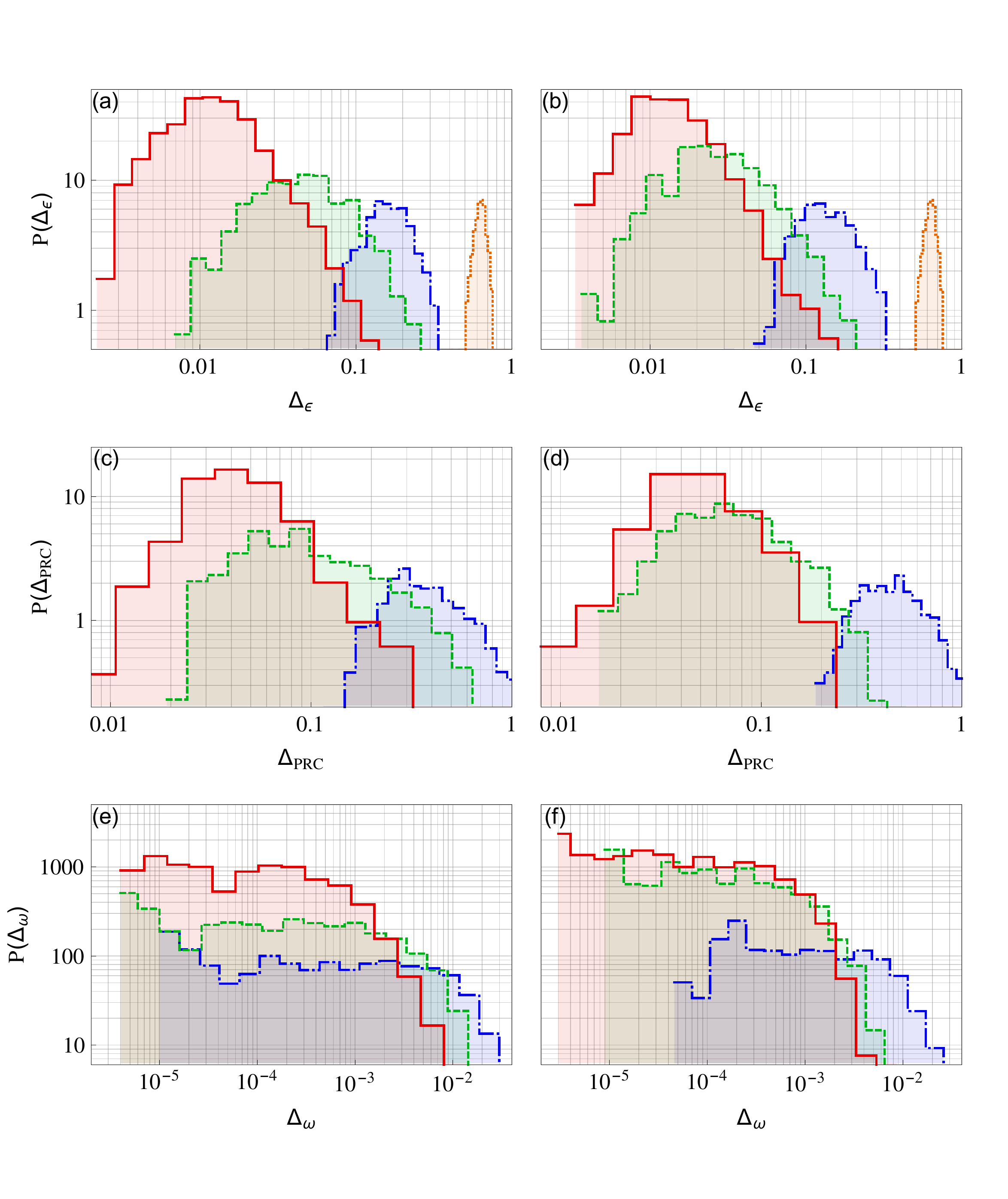}
\caption{(Color online) Histograms of errors of the 
coupling strengths $\Delta_{\e}$ (a,b),  PRC $\Delta_{\text{PRC}}$ (c,d), 
and frequency $\w$ (e,f), see Eqs.~(\ref{eq5}-\ref{eq7}).
Panels (a,c,e) and (b,d,f) correspond to tests with PRC type I and type II, 
respectively. 
In each panel the results of the first, third, and tenth iterations are 
shown in blue (dash-dotted), green (dashed), and red (solid line) respectively. 
In (a,b) also the distribution of errors for initial values $\e_i=1$ 
is shown in orange (dotted line).}
\label{fig:statanal}
\end{figure}

\begin{figure}[h!]
\centering
\includegraphics[width=1.01\columnwidth]{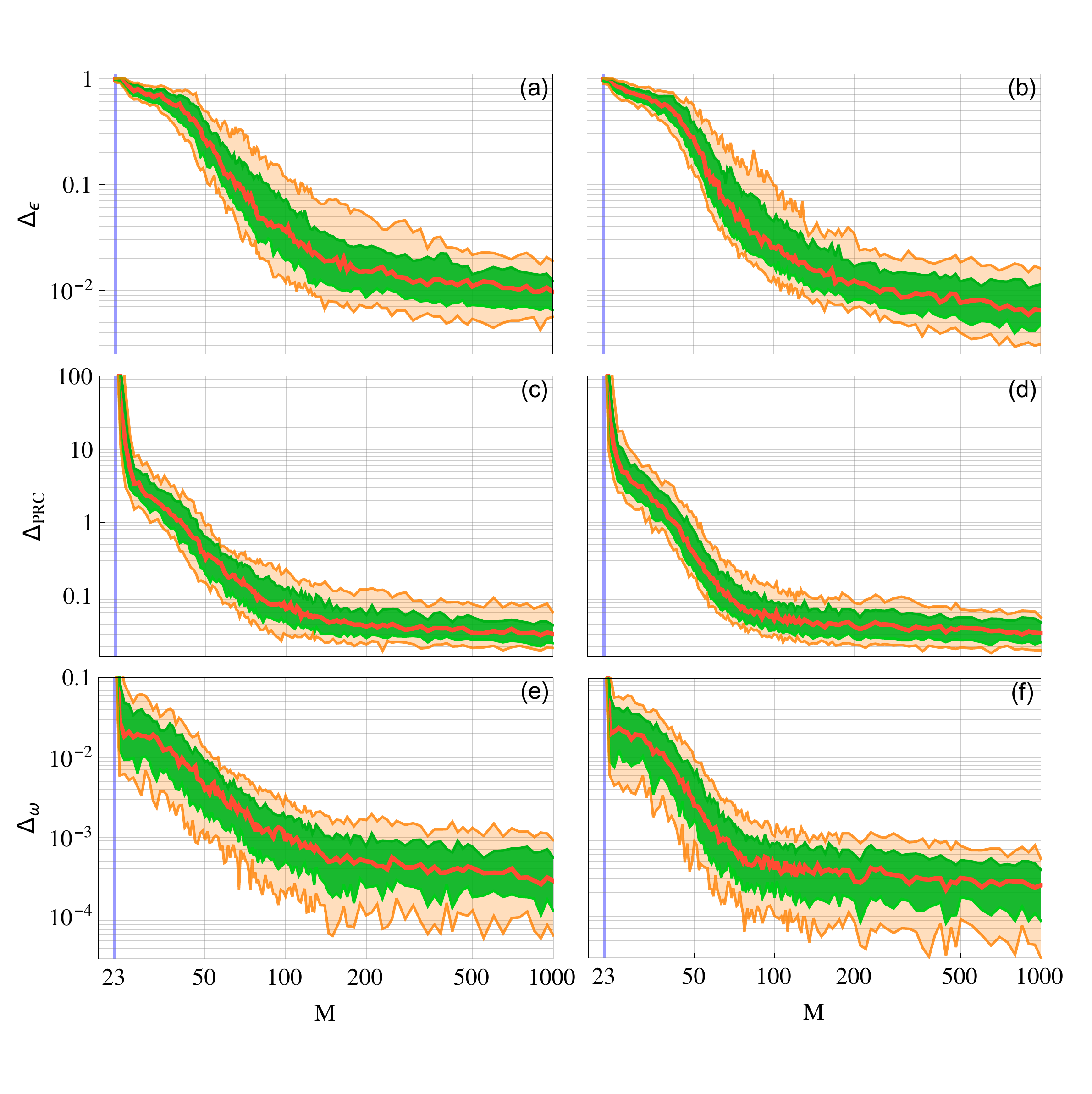}
\caption{(Color online) Error of the reconstruction in dependence on
the number of inter-spike intervals $M$ used for the analysis. 
Panels (a,c,e) and (b,d,f) show the results for PRC type I and PRC type II,
respectively. 
Panels (a,b) show the error of the coupling coefficients $\e_i$ (\ref{eq5}), 
panels (c,d) the error of the PRC (\ref{eq6}) and panels (e,f) the error of 
the frequency $\w$ (\ref{eq7}). 
For each value of $M$ reconstruction error was computed for 6000 different 
networks. 
The green (dark gray) and the orange (light gray) 
areas contain 50\% and 75\% of the errors, respectively;
the median is shown in red (bold line).}
\label{fig:errorM}
\end{figure}

Next, we perform a statistical analysis for $10^5$ network configurations. 
To quantify the 
quality of the reconstruction, we define the corresponding errors for 
recovered PRC, $\e_i$ and $\w$ as 
\begin{equation}
\Delta^2_{\text{PRC}} =  \frac{\int_0^{2\pi}\left[Z^{(\text{t})}(\vp)-Z^{(\text{r})}(\vp)
\right]^2 d\vp}{\int_0^{2\pi} \left[Z^{(\text{t})}(\vp) \right ]^2 d\vp}\;,
\label{eq5}
\end{equation}
\begin{equation}
\Delta^2_{\e} = \sum_{i=2}^{N}
\big[\e_i^{(\text{t})}-\e_i^{(\text{r})}\big]^2\bigg /
\sum_{i=2}^{N}\big[\e_i^{(\text{t})}\big]^2 \;,
\label{eq6}
\end{equation}
and
\begin{equation}
\Delta^2_{\w} = \big[\w^{(\text{t})}-\w^{(\text{r})}\big]^2\;.
\label{eq7}
\end{equation}
respectively~\footnote{Notice that error $\Delta_\e$ is minimized due to normalization 
according to Eq.~(\ref{eq:norm}).}.
The distributions of errors, shown in Fig.~\ref{fig:statanal}, confirm robustness of the 
iterative procedure.
Figure~\ref{fig:errorM} shows the dependence of the reconstruction error on the 
number of inter-spike intervals. 
Naturally, the more data we use, the better results we expect.
This test demonstrates, that reasonable reconstruction can be achieved 
already for several hundreds of intervals.

Finally, we performed the test with random assignment of the initial values
for the coupling coefficients $\e_i$. 
For several generated networks we performed $10^4$ reconstructions with 
different initial $\e_i$. The results confirm convergence of the algorithm 
for this case as well.

\subsection{Network of Morris-Lecar neurons}
\begin{figure}[h!]
\centering
\includegraphics[width=0.9\columnwidth]{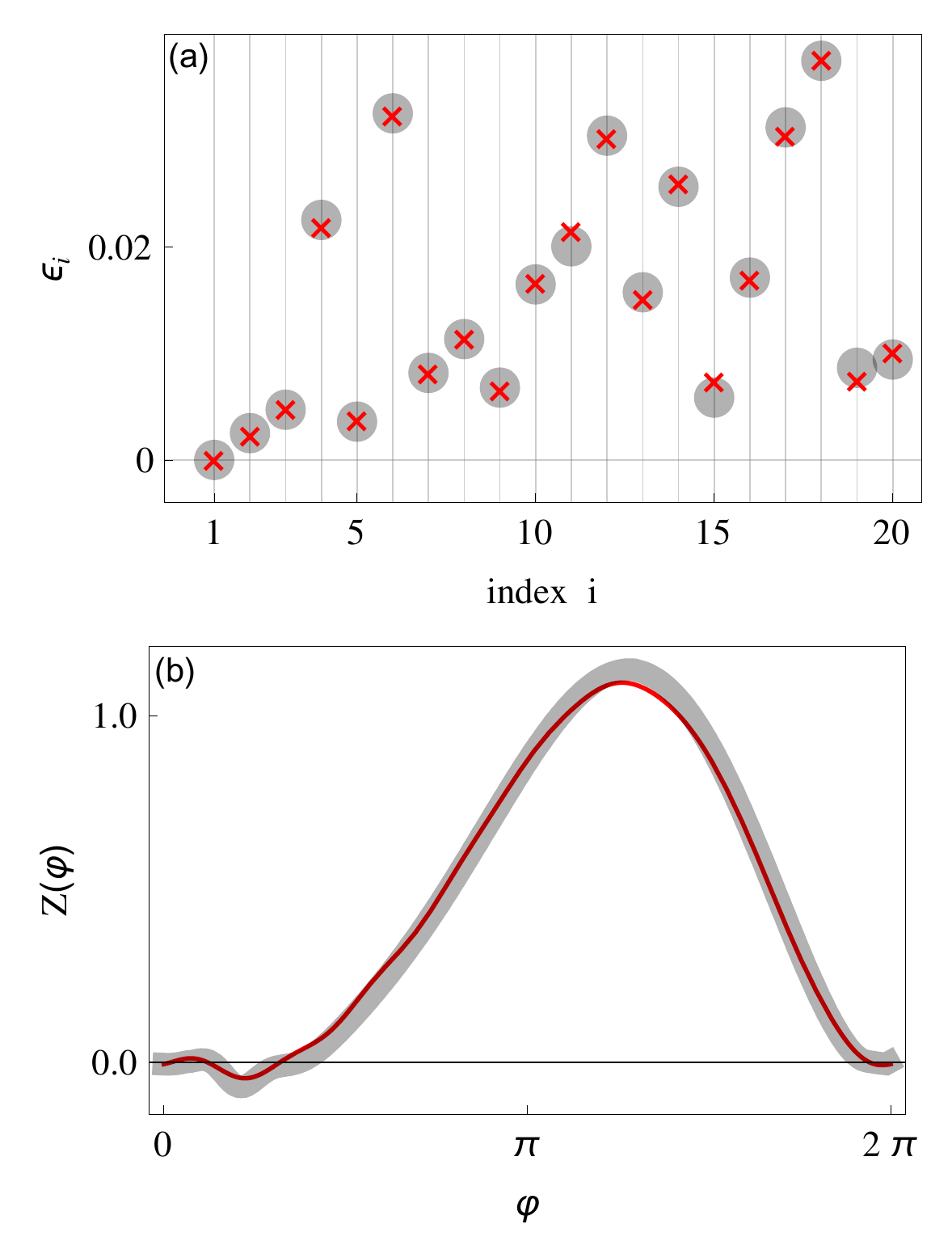}
\caption{(Color online) Reconstruction of a network of 20 Morris-Lecar neuronal 
oscillators, see Eqs.~(\ref{ml}-\ref{ml3}). 
200 inter-spike intervals were used for the reconstruction. 
(a) True values of the strength of the 
incoming connections for the first neuron (gray disks) 
and these values reconstructed after 10 iterations (red crosses).
(b) True (wide gray curve) and reconstructed (solid red curve) PRC. 
}
\label{fig:ml}
\end{figure}

In the next test we make a step towards more realistic modeling and consider a network 
of Morris-Lecar neurons \cite{Morris-Lecar-81,*Rinzel-Ermentrout-98}. The equations of the 
network are:
\begin{equation}
\begin{split}
\dot V_i=& I_i -g_l(V_i-V_l) -g_Kw_i(V_i-V_k)     \\
       & -g_{Ca}m_{\infty}(V_i)(V_{Ca}-V_i)+I_i^{\text{(syn)}}  \;, \\[1ex]
\dot w_i = &\lambda(V_i)(w_{\infty}(V_i)-w_i)\;,
\end{split}
\label{ml}
\end{equation}
where 
\begin{equation}
\begin{split}
m_{\infty}(V)&=[1+\tanh{(V-V_1/V_2)}]/2  \;,\\
w_{\infty}(V)&=[1+\tanh{(V-V_3/V_4)}]/2\;,\\
\lambda(V)&=\cosh{[(V-V_3)/2V_4]}/3\;,
\end{split}
\label{ml2}
\end{equation}
and $I_i^{\text{(syn)}}$ is the total incoming synaptic current.
We write the latter as
\begin{equation}
I_i^{\text{(syn)}}=\left[V_{\text{rev}}-V_i \right]\sum_{k,k\ne i}\frac{\e_{ik}}
{1+\exp{\left[-(V_i-V_{\text{th}})/\sigma\right]}}\;.
\label{ml3}
\end{equation}
 We take standard values for most of the parameters 
\footnote{Parameters of the system (\ref{ml},\ref{ml2}) are:
$g_L=0.5$, $g_K=2$, $V_1=-0.01$, $V_2=0.15$, $V_{Ca}=1$, $V_K=-0.7$, 
$V_L=-0.5$, $g_{Ca}=1.33$, $V_3=0.1$, $V_4=0.145$.
}.
Parameters of the synaptic coupling are $V_{\text{rev}} = 0.2$, 
$V_{\text{th}}=0.25$, and $\sigma=0.01$.
The neurons are non-identical: the values of the current are 
$I_i=0.077(1+0.22\xi)$, where $\xi$ is uniformly distributed between 
zero and one; for these values the neurons remain in the spiking state.
The results of the analysis with $200$ inter-spike intervals 
shown in Fig.~\ref{fig:ml}, confirm efficiency of our technique.

\section{Discussion and conclusions}
\label{sec:concl}

With the help of two model systems we have demonstrated, that 
our technique provides a robust reconstruction of a network. 
The data requirements are not too demanding:
the reconstruction is quite precise already for time series 
of several hundreds of spikes.
Now we discuss some limitations of the method.

First we comment on the initial estimate of phases 
using Eq.~(\ref{eq3}). As already mentioned, 
the error is proportional to $\e_i\Vert Z\Vert$
and increases with the number of spikes that arrive 
within the inter-spike interval of the driven unit.
This explains why the case of the slowest oscillator is 
the most difficult one: such an oscillator has on average more 
incoming stimuli per inter-spike interval then the fast units. 
This means, that though our examples demonstrate robust 
reconstruction, it may fail if $\w_i/\w_1\gg 1$.

Next limitation is related to variability of the inter-spike 
intervals of the driving unit $i$ (for $\w_i>\w_1$). 
Indeed, suppose that drive is strictly periodic. Then time of 
the appearance of the first spike unambiguously determines
the timing of the following ones, and hence, the length of 
the inter-spike interval $T_k$. 
However, $T_k$ is then determined by the sum of different pieces of PRC
and this sum cannot be disentangled.  
The initial estimation of the strength of the connection 
as described in Appendix~\ref{subsec:first_estimate}
can still work, but the recovery of the PRC becomes impossible
and the iterative procedure fails.
So, we foresee that reconstruction may be not so robust for 
very sparse network where we expect to have purely periodic nodes. 
On the other hand, a realistic network 
is noisy, and noise naturally provides the desired 
variability in the time series, thus enhancing the reconstruction.
Finally, we mention that the reconstruction fails if the network 
synchronizes.

\section*{Acknowledgment}
We  acknowledge useful discussions with A. Pikovsky, M. Zochowski, 
R. Andrzejak, and A. Daffertshofer.
This work has been financially supported by the EU project COSMOS (642563).

\appendix
\section{First estimation of incoming connections}
\label{subsec:first_estimate}
For sufficiently long data an initial estimation of the coupling 
strength $\e_i$ can be performed by evaluating the effect of the first 
pulse from the unit $i$, that arrives
within the $k$-th inter-spike interval, on the length of this interval $T_k$.
For this purpose, we first plot $T_k$ vs $\vp_k^{(i,1)}$, for all incoming links.
Next, for each plot, we divide the $\vp$-axis into $N_b$ bins and average 
the $T_k$ values within each bin. As a result, we obtain a dependence 
$\bar T^{(i)}_n\left(\bar\vp_{n}\right)$, where 
$\bar\vp_{n}=\frac{\pi}{N_b}(2n-1)$, $n=1,\ldots,N_b$, 
are phases at the centers of bins. Our conjecture is that
$\bar T^{(i)}_n\left(\bar\vp_{n}\right)$ reflects the strength 
of the incoming connection: if this strength is zero, i.e. there is no 
incoming link from unit $i$, then there shall be no dependence; 
on the other hand, if the incoming connection is strong, then we expect 
the dependence to be well-pronounced.

\begin{figure}[h!]
\centering
\includegraphics[width=0.98\columnwidth]{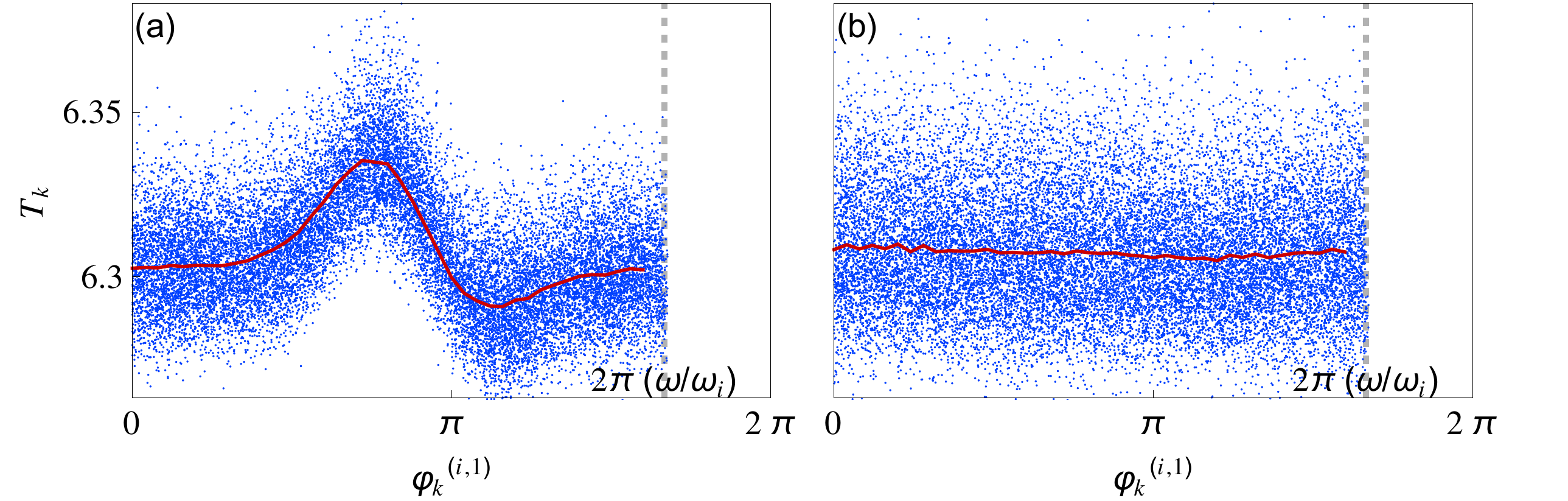}
\caption{(Color online) Scatter plots of inter-spike intervals $T_k$ vs.
approximated phase $\vp_k^{(i,1)}$ of the first spike from a chosen
driving oscillator $i$, for a strong coupling strength (a) and for 
a weak one (b). The horizontal axis is divided into $N_b=50$ bins and
solid red curve shows the average of $T_k$ over each of 
the bins, $\bar T^{(i)}$, as a function of the central phase $\bar \vp$ 
of the bins; $2\cdot10^4$ inter-spike 
intervals are used in this computation.
}
\label{scatters}
\end{figure}

We illustrate this idea in Fig.~\ref{scatters}, where two such plots are 
shown for the cases of weak and strong incoming connections. 
We see that indeed $\bar T^{(i)}_n\left(\bar\vp_{n}\right)$ reflects
the coupling coefficients $\e_i$. Hence, we use the standard deviation of 
this dependence as the first estimate, i.e. we take
\begin{equation}
\e_i=\Big\langle \big(\bar T^{(i)}_n -\langle \bar T^{(i)}_n 
\rangle\big)^2 \Big\rangle^{1/2} \;,
\label{eq4}
\end{equation}
where $\langle \cdot \rangle$ means averaging over $N_b$ bins. 
Although the correspondence between the estimate and true coupling strength 
is not exact, in most cases this approach yields reasonable values.

%

\end{document}